\documentclass[aps,prl,showpacs,amsmath,amssymb,
	preprint,onecolumn]{revtex4}

\usepackage[dvips]{graphicx}
\usepackage{dcolumn}
\usepackage{bm}
\usepackage{epsf}

\begin{document}

\title{Predictions of 
ultra-harmonic oscillations in coupled arrays of limit cycle
  oscillators}

\author{Alexandra S. Landsman$^1$, Ira B. Schwartz$^1$}
\affiliation{$^1$US Naval Research Laboratory, Code 6792, Nonlinear Systems Dynamics Section, Plasma Physics Division, Washington, DC 20375}
\email{<alandsma@cantor.nrl.navy.mil>}

\begin{abstract}
Coupled distinct arrays of nonlinear oscillators have been shown to have 
a regime of high frequency, or ultra-harmonic, oscillations that are at
 multiples of the natural frequency of individual oscillators. The
 coupled array architectures generate an in-phase high-frequency
 state by coupling with an array in an anti-phase state. 
The underlying mechanism
 for the creation and stability of the ultra-harmonic oscillations is
 analyzed.  A class of inter-array coupling is shown to create a
 stable,  in-phase oscillation having frequency that increases linearly
 with the number of oscillators, but with an amplitude that stays
 fairly constant.  The analysis of the theory is illustrated by
 numerical simulation of coupled arrays of Stuart-Landau limit cycle
 oscillators. 
\end{abstract}


\maketitle

\section{\label{sec:intro} I. Introduction}

The dynamics occurring in coupled oscillators has been of much interest
in both basic and applied science. Since the original early observations
of Huygen's of two synchronized coupled nonlinear clocks (See the
appendix for a nice description in \cite{pikovsky}), theories of
coherent motion in limit cycle and phase oscillator arrays having
different couplings have become numerous \cite{Daido92,Daido93,KuramotoN87,Strogatz00}.
In most cases studied, however, the arrays are assumed to have a single
statistical coupling architecture, such as global coupling \cite{PikovskyPM01,TeramaeK01},
nearest neighbor and ring coupling \cite{BuonoGP00,El-NasharZCI03},
and long range coupling \cite{MarodidV02}, to list just a few. 

In many of the examples of coherent motion studied, 
the synchronized in-phase state and incoherent splay or anti-phase
state have both been observed. In coupled semi-conductor lasers, it
is known that nature prefers the incoherent states \cite{WinfulW88} , while in Josephson
junction arrays which are globally coupled, both in-phase and incoherent
states may co-exist \cite{SchwartzT94,TsangS92}. The stability of
the in-phase state in most applications has been important in order to maintain frequency
control and/or power output in the arrays, such as in coupled lasers and
Josephson's junctions.

Recently, interacting arrays of limit cycle oscillators possessing
different architectures have been proposed to stabilize a high frequency
in-phase state \cite{PalaciosCLRIKNMB05} .  While numerical simulations
exhibit stable in-phase states over a particular range of parameters, 
no analytical work has been done on the system,
and the mechanism behind the onset of stable 
ultraharmonic oscillations has not been explained.  
The goal of this paper is to explain the mechanism behind the onset of stable
ultraharmonics, 
obtain analytic results for the critical value of the coupling constant
required for this onset, 
as well as to explore an alternative
coupling scheme previously not considered.
The paper is organized as follows:  In Section II, the equations
of motion are introduced and solved exactly, in the absence of global coupling
between the two arrays.  In Section III, the bifurcation value of the global
coupling constant above which stable ultra-harmonic oscillations set in is
calculated and shown to depend of frequency, amplitude and total number of
oscillators.  The physical mechanism behind the onset of stable
ultra-harmonics is explained, as resulting from a large constant component in
the amplitude-dependent global coupling, that determines the bifurcation
value.  Above the bifurcation value, the dynamics of one of the arrays become
a function of the other, resulting in ultraharmonics.
Averaging theory is applied in Section IV to explain the equivalence of the
bifurcation diagrams between the averaged and the full system and to estimate
the amplitude of ultraharmonic oscillations.  Section V derives an
alternative type of coupling that while more complicated in form, has the
advantage of creating a higher amplitude ultra-harmonic oscillations, with an
amplitude that stays fairly constant as the number of oscillators in the array
increases.  Section VI concludes and summarizes our results.

\section{\label{sec:basic} II. Basic Equations and Dynamics}

\begin{figure}[ht]
\hspace*{-1 cm}
{
\epsfxsize=6in
\epsffile{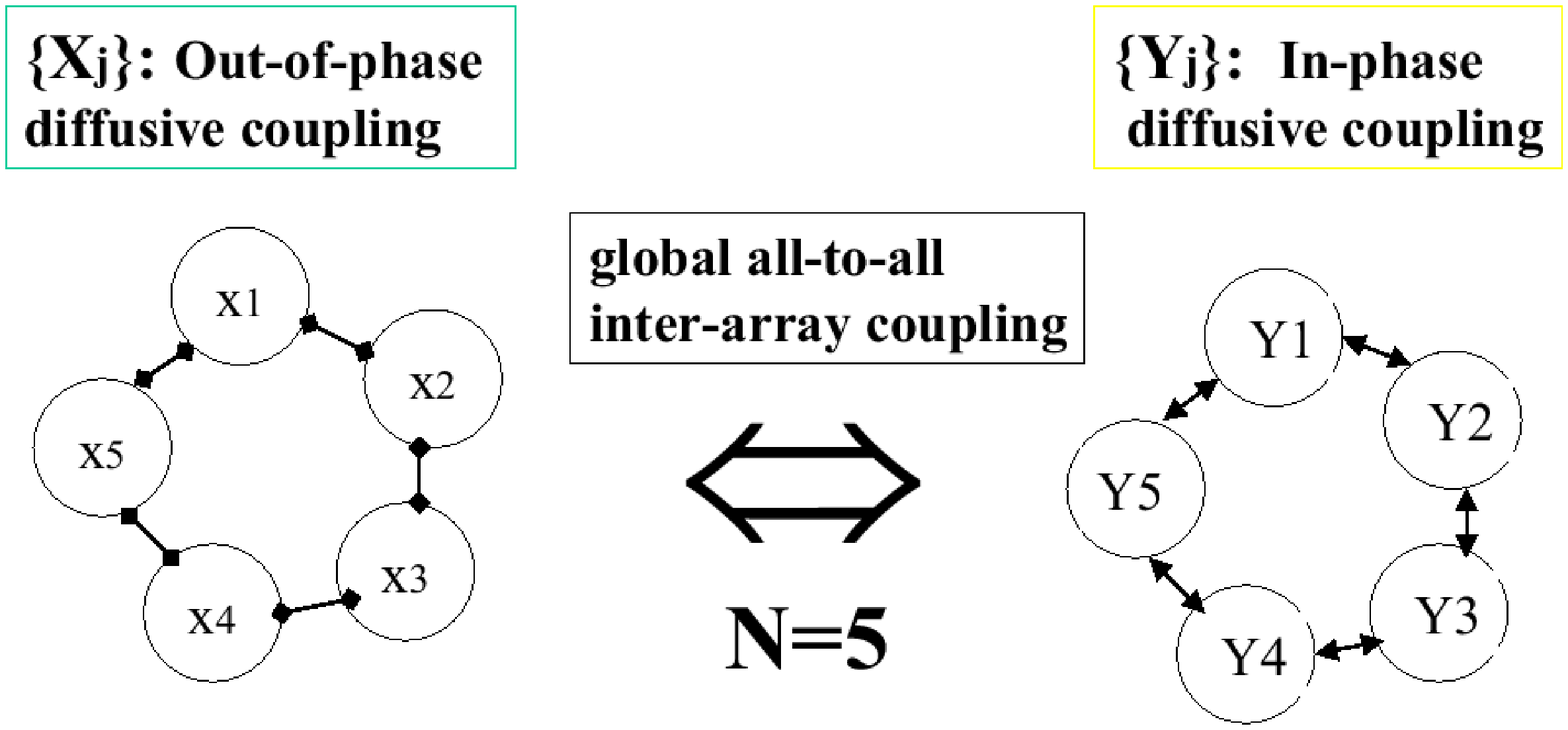}
}
\caption{Two diffusively coupled arrays coupled to each other via
global coupling, for $N=5$.  The $X$ array has out-of-phase diffusive
coupling, and the $Y$ array has in-phase diffusive coupling.  The two
arrays are globally coupled to each other.}
\label{fig:2}
\end{figure}

The internal dynamics of the uncoupled 
Stuart-Landau oscillators in normal form is:
\begin{equation}
\dot{\vec{z}} = (\alpha + i \omega) \vec z - |\vec z|^2 \vec z 
\label{eq:z}
\end{equation}
where $\vec z$ is a complex variable. 
Equation (\ref{eq:z}) can be easily solved by transforming into polar
variables, and results in a limit cycle oscillator 
with frequency equal to $\omega$ and steady-state
amplitude of oscillation equal to $\sqrt{\alpha}$.
The system considered here consists of two arrays of coupled Stuart-Landau
oscillators, $\{X_j\}$ and $\{Y_j\}$,
where the index, $j$ stands for each individual oscillator.  
Throughout the rest of the paper, the $\{X_j\}$ and $\{Y_j\}$ will be referred
to simply as $X$ and $Y$ arrays.
In the absence of
coupling, the equations of motion of each oscillator within the arrays is given
by Eq. (\ref{eq:z}).  We start by considering two coupled arrays, as initially
introduced in  \cite{PalaciosCLRIKNMB05}.
The difference
between the two arrays is that $X$ 
oscillators possess anti-phase diffusive coupling
between oscillators within the array, 
while the $Y$ array has in-phase diffusive coupling within the array.
The two arrays are globally coupled to each other.  The whole system is
described by the following set of equations:
\begin{equation}
\dot{\vec{X_j}} = (\alpha_x + i \omega) \vec X_j - |\vec X_j|^2 \vec X_j +
c_x\left(X_{j-1} + X_{j+1} - 2 X_j\right) + c_{yx} \sum_{k=1}^N |Y_{k}|
\label{eq:xr}
\end{equation}

\begin{equation}
\dot{\vec{Y_j}} = (\alpha_y + i \omega) \vec Y_j - |\vec Y_j|^2 \vec Y_j +
c_y\left(Y_{j-1} +  Y_{j+1} - 2 Y_j\right) + c_{xy} \sum_{k=1}^N |X_{k}|
\label{eq:yr}
\end{equation} 
A schematic diagram of the
architecture of the coupled arrays in Eqs. (\ref{eq:xr}) and (\ref{eq:yr}) is
shown in Figure (\ref{fig:2}).
In the case of the $X$ array oscillators,
the diffusive coupling constant of $c_x < 0$ 
between oscillators within the array leads to an anti-phase solution.
For the $Y$-array oscillators, where $c_y > 0$, 
the in-phase state is the only stable solution.  
Thus in steady-state, all oscillators in the $Y$-array tend
to have the same dynamics, so that the diffusive coupling term in
Eq. (\ref{eq:yr}) drops out.
The $Y$-array oscillators will all oscillate in-phase in steady-state, which
is a direct outcome of the diffusive coupling term, $c_y > 0$, 
so that the in-phase state is the only stable state.  
From Eqs. (\ref{eq:xr}) and (\ref{eq:yr}), 
it is clear that global coupling depends only on the
 amplitude of oscillators in the two arrays, $|Y_{k}|$ and $|X_{k}|$.  If
 this amplitude is approximately conserved, then the main contribution from the
 global coupling term will be a constant component.  In Section V, an
 alternative coupling scheme that results in a higher oscillatory component in
 the global coupling, leading to higher intensity ultraharmonics, 
will be considered.

\subsection{\label{sec:subsec1} Dynamics in the absence of inter-array
  coupling}
In the absence of inter-array couping, $c_{xy}=c_{yx}=0$, an exact
steady-state solution can be obtained for the two arrays.  
Since in steady-state, all the $Y$-array oscillators move in-phase,
the diffusive
coupling term drops out of the Eq. (\ref{eq:yr}), and the solution reduces
to that of a single limit cycle oscillator, so that the index $j$ on the $Y$
array will be
dropped throughout the rest of the paper.  
Unlike the $Y$ array oscillators, the diffusive coupling term
in the $X$ array is negative, so that the oscillators tend to be
maximally out-of-phase with their nearest neighbors.  For a ring of
diffusively coupled oscillators, this results in two solutions, depending on
whether $N$ is even or odd:
\begin{equation}
\triangle \phi_{j,j+1}^{even} = \pi
\label{eq:Neven}
\end{equation}
\begin{equation}
\triangle \phi_{j,j+1}^{odd} = \pi + \pi/N
\label{eq:Nodd}
\end{equation}
where $\triangle \phi_{j,j+1}^{even}$ and $\triangle \phi_{j,j+1}^{odd}$ is
the phase difference between the $j$th and the $j+1$th 
oscillator in the $X$-array for $N$-even and $N$-odd, respectively.
Since all the oscillators in the $X$-array are identical and differ only by a
phase-shift, we now have an expression for the 
steady-state dynamics of $j+1$ and
$j-1$ oscillators as a function of the $j$th oscillator:
\begin{equation}
\vec X_{j-1} = e^{-i \triangle \phi_{j,j+1}} X_j
\qquad \vec X_{j+1} = e^{i \triangle \phi_{j,j+1}} X_j
\label{eq:phase}
\end{equation}
Substituting Eq. (\ref{eq:phase}) into Eq. (\ref{eq:xr}), using Eqs.
(\ref{eq:Neven}) and (\ref{eq:Nodd}), and 
setting $c_{yx}=0$, we can now solve for
the amplitude of the $X$-array oscillators in the absence of global
inter-array coupling: 
\begin{equation}
A_x^{N-even} = \left(\alpha_x - 4 c_x\right)^{1/2}
\label{eq:AmpEven}
\end{equation}
for even $N$, and
\begin{equation}
A_x^{N-odd} 
= \lbrack \alpha_x - 2 c_x \left( 1 + cos(\pi/N)\right) \rbrack ^ {1/2}
\label{eq:AmpOdd}
\end{equation}
for odd $N$.
The frequency of $X$ array oscillators is the same as in the $Y$ array,
the limit cycle frequency, $\omega$.  For $\alpha_x = \alpha_y$
however, since $c_x < 0$, 
the amplitude of
oscillation is higher than that of the in-phase $Y$-array, 
which oscillates at
amplitude $\alpha_y^{1/2}$.  As will be shown in the following section, this
difference in amplitude is important to the creation of the stable
ultraharmonics and affects the range of the global coupling constant
necessary for stable ultraharmonics, in
the case of symmetric coupling.

From Eqs. (\ref{eq:Neven}) and (\ref{eq:phase}), it is clear that for
$N$-even, the $X$ array splits into two identical subgroups containing $N/2$
oscillators, where the two subgroups are out-of-phase with each 
other by $\pi$. For $N$-odd, the $X$-array forms a traveling wave, with phases
distributed over the unit circle in increments of $2 \pi/N$.  This can be seen
by using Eq. (\ref{eq:Nodd}), which gives $2 \pi/N$ as the smallest phase
difference (between $j$th and $j+2$nd oscillators).  Since for generation of
ultraharmonics at frequency $N \omega$, the $X$ array must form a traveling
wave, with phase increments of $2 \pi/N$, the rest
of the paper will focus on $N$-odd arrays.  The reason
why this  $2 \pi/N$ distribution of phases is needed for the creation of
ultraharmonics at frequency $N \omega$ will be explained in the following
section.  Other types of inter-array coupling schemes could be considered
that would result in a different phase distribution.

\section{\label{sec:interact} III. The effect of global coupling and 
creation of stable ultraharmonics via a bifurcation}

It was shown in the previous section that 
in the absence of global coupling, $c_{yx}=c_{xy}=0$, 
the steady state dynamics of the two arrays are that of a simple harmonic
oscillator, with amplitudes given by $\alpha_y^{1/2}$ and 
Eq. (\ref{eq:AmpOdd}),
for the $Y$ and $X$ arrays, respectively.  Since the global coupling term is a
function of amplitude only (see Eqs. (\ref{eq:xr}) and (\ref{eq:yr})), we
would expect the global coupling to introduce a constant component into
Eqs. (\ref{eq:xr}) and (\ref{eq:yr}).  In fact, if the coupling is one-way,
that is if we set $c_{yx}=0$, then the global coupling
term only contributes a constant component and no ultraharmonics can occur
in the $Y$ array.  This explains the previously made observation
\cite{PalaciosCLRIKNMB05} that mutual coupling between the two arrays is
required to induce ultraharmonic oscillations.  It follows that the 
global coupling term in the $X$ array must create a perturbation in the
amplitude of the $X$ oscillators, 
so that it is no longer a conserved quantity. 
To understand how this happens, we examine the weakly coupled case,
$c_{yx}, c_{xy} \ll 1/\left(\alpha_{x,y}^{1/2} N\right)$.  
Then, in the absence of resonant interactions, the
steady-state dynamics of the two arrays is given to lowest order by:
\begin{equation}
\dot{\vec{X_j}} = (\tilde \alpha_x + i \omega) \vec X_j - |\vec X_j|^2 \vec X_j 
+ C_y
\label{eq:xrC}
\end{equation}

\begin{equation}
\dot{\vec{Y}} = (\alpha_y + i \omega) \vec Y - |\vec Y|^2 \vec Y + C_x
\label{eq:yrC}
\end{equation}
where $\tilde \alpha_x =  \alpha_x - 2 c_x \left( 1 + cos(\pi/N)\right)$.
The index $j$ has been dropped in the $Y$ array, since the array oscillates
in-phase in steady-state.  The subscript on $C_y$ ($C_x$) indicates
that the term comes from the global coupling from the $Y$ to the $X$ array
(from the $X$ to the $Y$ array).  
The constant components $C_x$ and $C_y$ in Eqs. (\ref{eq:xrC}) and
(\ref{eq:yrC}) are given by:
\begin{equation}
C_y = c_{xy} N \alpha_y^{1/2}; \qquad C_x = c_{xy} N \lbrack \alpha_x 
- 2 c_x \left( 1 + cos(\pi/N)\right) \rbrack ^{1/2}.
\label{eq:CxOdd}
\end{equation}
Since $c_x < 0$, $C_x > C_y$,  it is  clear that the $X$ array
has a stronger effect on the dynamics of the $Y$ array than vice versa, even
in the case of symmetric coupling: $c_{xy}=c_{yx}$.  
Equations of the form (\ref{eq:xrC}) and (\ref{eq:yrC}) have been studied
previously in connection with forced Van der Pohl oscillators
\cite{guckenheimer}.  The main characteristic of these equations is that they
execute a limit cycle below a certain critical value of the constants, $C_y$,
$C_x$.  Above the critical bifurcation value, the oscillations are damped out,
so that the system reaches a stable equilibrium (a sink).  There is also a
narrow intermediate range (between the limit cycle and the sink region) 
which will
not be discussed, but for a more detailed description see \cite{guckenheimer}.
Equation (\ref{eq:xrC}) can be used to understand how ultraharmonic
oscillations are created in the coupling term, $c_{xy} \sum_{k=1}^N |X_{k}|$
in Eq. (\ref{eq:yr}).  The constant component, $C_y$, of the global coupling
in Eq. (\ref{eq:xrC}), breaks the symmetry about the origin,
making the amplitude of oscillation
dependent on the phase of the oscillator.  
In phase-space, this corresponds to the shifting of
the center of limit cycle from the origin.  Now, all of the oscillators in the
$X$ array have amplitudes which oscillate at the frequency of the limit cycle,
$\omega$.  Since the phases of the oscillators in the $X$ array are
distrubuted in increments of $2 \pi/N$, there are $N$ peaks in amplitude over
a single limit cycle oscillation.  Adding 
all of the amplitudes together in the coupling term, $c_{xy} \sum_{k=1}^N
|X_{k}|$ creates an ultraharmonic oscillation at frequency $N \omega$.  
It follows that the global coupling term in the $Y$ array will have a
component oscillating at an ultraharmonic frequency, $N \omega$ as well as a
much larger constant component, $C_x$.  If this constant component, $C_x$, is
above the bifurcation value, $C_x > C_{bx}$, the $Y$ array will be driven into
the steady-state region of phase-space.  It will then execute ultraharmonic
oscillations with frequency $N \omega$ about that steady state.  This can be
better explained using averaging theory that applies when the system,
given by Eq. (\ref{eq:yrC}) is close to steady-state (when  $C_x > C_{bx}$)
and will be explored in the following section.

The bifurcation values $C_{bx}$ and $C_{by}$ for $C_x$ and $C_y$, respectively,
occur when the the real part of the
nullcline straightens out, leading to a single
equilibrium, which is a steady state.  Calculating the bifurcation value for
Eqs. (\ref{eq:xrC}) and (\ref{eq:yrC}),
\begin{equation}
C_{b} = 2\left(\frac{\tilde \alpha}{3}\right)^{3/2} + \left(\frac{3}{16
 \tilde \alpha}\right)^{1/2} \omega^2
\label{eq:Cb}
\end{equation}
where $\tilde \alpha = \alpha_y$ for $C_{bx}$ and 
$\tilde \alpha = \alpha_x  - 2
c_x \left( 1 + cos(\pi/N)\right)$ for $C_{by}$.  For $\tilde \alpha/\omega
\geq 1$, $C_b$ is a monotonically increasing function of $\tilde \alpha$, so
that the critical bifurcation for the $X$ array is higher then for the $Y$
array (for $\alpha_x = \alpha_y$), since $c_x < 0$.   
 We are now in a position to give an
explanation of the mechanism behind 
the onset of ultraharmonic oscillations in the $Y$ array.  As
the symmetric global coupling constant, $c_{xy}=c_{yx}$, increases, 
a critical
bifurcation value is reached in the constant component of the global coupling,
whereby the $Y$ array undergoes a bifurcation into a sink region of the
phase-space, while the $X$ array is still executing a limit cycle
oscillation.  
After undergoing a bifurcation, the $Y$ array oscillators
are driven by the oscillatory
ultraharmonic component of the coupling from the
$X$-array and execute ultraharmonic oscillations about a steady state, which
is determined by the constant component, $C_x$.

\begin{figure}[ht]
\hspace*{-1 cm}
{
\epsfxsize=6in
\epsffile{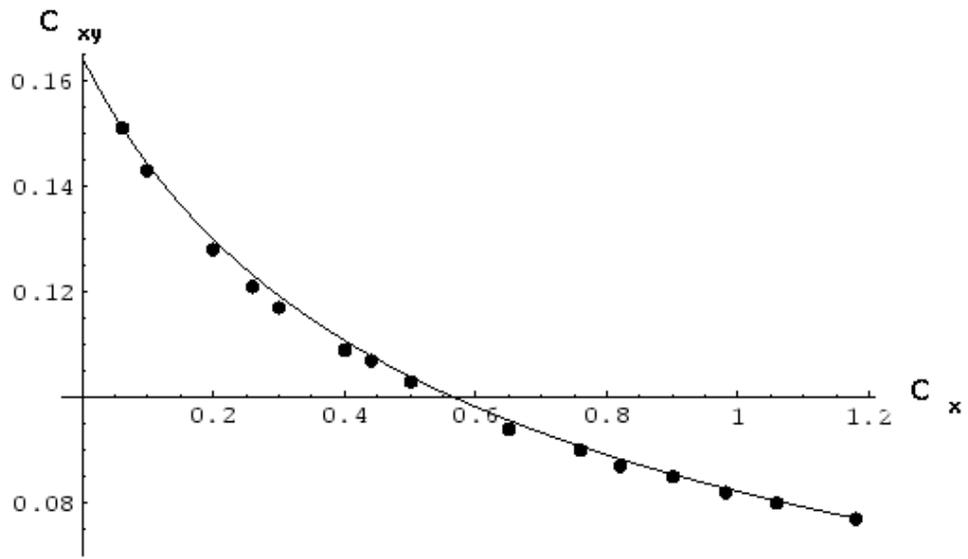}
}
\caption{Bifurcation values of the global coupling constant, $c_{xy}$ as a
function of out-of-phase diffusive coupling, $c_x$.  The solid line is a plot
of the analytically derived Eq. (\ref{eq:Cb_Fsolve}).  The numerical values are
also shown on the graph}
\label{fig:3}
\end{figure}

The constant component, $C_x$, of the coupling term  $c_{xy} \sum_{k=1}^N
|X_{k}|$ in Eq. (\ref{eq:yr}) is given by 
\begin{equation}
C_x = c_{xy} N \bar A_x
\label{eq:cxSimple}
\end{equation}
where $\bar A_x$ is the average amplitude of the $X$ array.
If the amplitude of the $X$ array is not substantially affected by the global
coupling term, then $\bar A_x$ can be approximated by Eq. (\ref{eq:AmpOdd}).
Substituting $C_{bx}$ from Eq. (\ref{eq:Cb}) 
and $A_x$ from Eq. (\ref{eq:AmpOdd}) into Eq. (\ref{eq:cxSimple}) and solving
for $c_{xy}$,  we get an approximation of the
bifurcation value of the global coupling constant
for the onset of stable ultraharmonic oscillations,
\begin{equation}
c_{xy}^b =  \frac{2 \left(\alpha_y/3\right)^{3/2} + \left(3/16 \alpha_y\right)^{1/2}
\omega^2 }{N \lbrack \alpha_x - 2 c_x \left( 1 + cos(\pi/N)\right) \rbrack ^ {1/2}}
\label{eq:Cb_Fsolve}
\end{equation}
Figure (\ref{fig:3}) plots the bifurcation values of the global
coupling constant, given by Eq. (\ref{eq:Cb_Fsolve}) as a function of the
diffusive coupling in the $X$ array, $c_x$.  Note the excellent agreement
between the analytically derived curve and the numerical simulation.  The
slight consistent under-estimate of $c_{xy}^b$ comes from using the
unperturbed amplitude, $A_x$, as given by Eq. (\ref{eq:AmpOdd}).  As can be
seen from the figure, however, the unperturbed amplitude, $A_x$ is an
excellent approximation for the globally coupled system over a whole range of
values.  

\begin{figure}[ht]
\hspace*{-1 cm}
{
\epsfxsize=7in
\epsffile{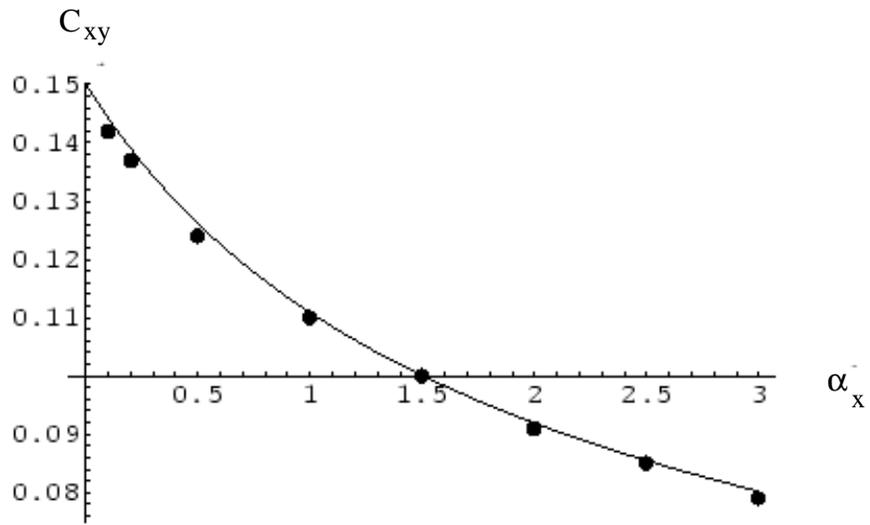}
}
\caption{Bifurcation values of the global coupling constant, $c_{xy}$ as a
function of amplitude, $\alpha_x$.  
Both the analytically derived dependence and
the numerics are shown}
\label{fig:3b}
\end{figure}

Figure (\ref{fig:3b}) plots the bifurcation values of the global
coupling constant, given by Eq. (\ref{eq:Cb_Fsolve}) as a function of the
parameter, $\alpha_x$.  Higher values of $\alpha_x$, relative to $\alpha_y$
lead to higher relative amplitude of oscillation of the $X$ array, resulting
in a lower bifurcation value of $c_{xy}^b$ and a greater range of the global
coupling constant,  $c_{xy}^b$ for generation of ultraharmonics.
From Eq. (\ref{eq:Cb_Fsolve}), for 
large values of the diffusive coupling, $c_x$, the bifurcation value for
the onset of ultraharmonic oscillations decreases as $1/\sqrt{|c_x|}$. 
Since Eq. (\ref{eq:Cb}) is a monotonically increasing function of  $\tilde
\alpha$, it is clear that the bifurcation value of the $X$ array into a sink
increases as a function of $|c_x|$.  It follows that as $|c_x|$ increases, the
range of the coupling constant $c_{xy}=c_{yx}$ whereby stable ultraharmonics
are induced in the $Y$-array also increases.  If the value of the global
coupling is too high, then the $X$ array undergoes a bifurcation, resulting in
oscillation death.  So, the coupling  constant should be
high enough, $c_{xy} > c_{xy}^b$ so that
the dynamics in Eq. (\ref{eq:yrC}) bifurcate from limit cycle into a sink, 
but low enough so that the oscillators in the $X$ array undergo a limit
cycle oscillation at the natural frequency, $\omega$.

\section{\label{sec:averaging} IV. Averaging theory and the creation of
  ultraharmonic oscillations}
\begin{figure}[ht]
\hspace*{-1 cm}
{
\epsfxsize=6in
\epsffile{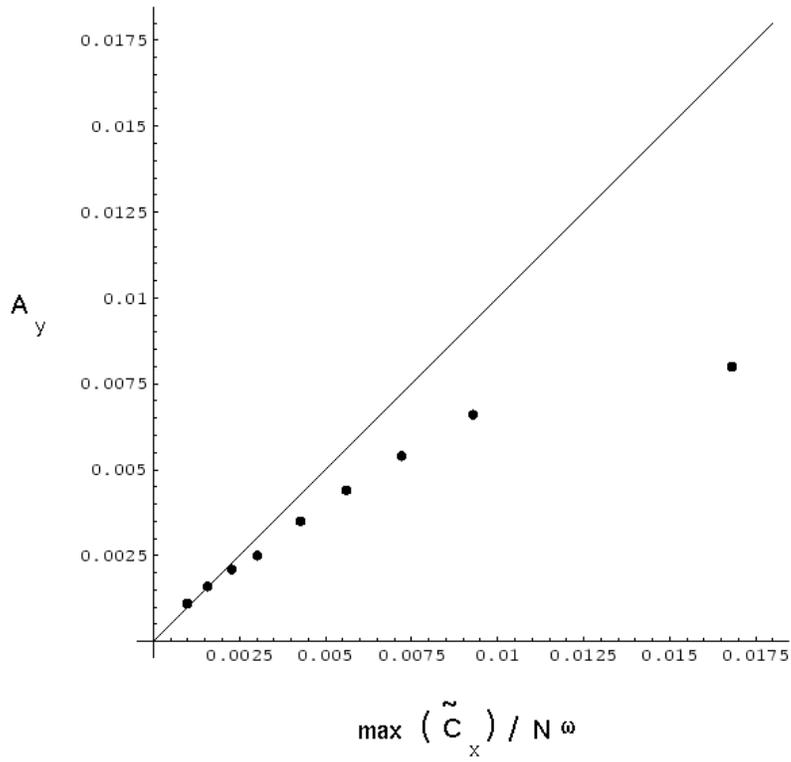}
}
\caption{Amplitude of ultraharmonic oscillations, $A_y$ vs. 
$max(\tilde C_x)/N \omega$.
At lower amplitudes, the oscillations fall on the $A_y = max(\tilde C_x)/N
\omega$ line, in accordance with Eq. (\ref{eq:average2}). }
\label{fig:ave}
\end{figure}

The mechanism behind the onset of ultraharmonics has been explained in the
previous section.  For small amplitude ultraharmonics, averaging theory can
be used to prove that the frequency of these
oscillations is the multiple of the limit-cycle frequency, $N\omega$.  
Averaging is applicable to the systems of the form 
\begin{equation}
\dot x = \epsilon f\left(x,t\right), \qquad x \in \mathcal{R}^n
\label{eq:averaging}
\end{equation}
where $\epsilon$ is small and $f$ is, $T$-periodic in $t$
\cite{guckenheimer}.  In our case, the periodic forcing comes from the
ultraharmonic frequency, $N \omega$, in the periodic coupling term  $c_{xy}
\sum_{k=1}^N |X_{k}|$.  In system of the form given by
Eq. (\ref{eq:averaging}), the relatively high frequency of periodic forcing
constrasts with the slow evolution of the averaged system.  Thus the averaging
theory can be applied if the amplitude of ultraharmonic oscillations about the
equilibrium given by solving Eq. (\ref{eq:yrC}) is small.   
As previously explained, the global coupling term can be broken up into a
large constant component, $C_x$, and the relatively small 
oscillatory component, $\tilde C_x$:
\begin{equation}
c_{xy} \sum_{k=1}^N |X_{k}| = C_x + \tilde C_{x}(t)
\label{eq:CouplingSplit}
\end{equation}
where $\tilde C_x \ll C_x$. 
Dividing Eq. (\ref{eq:yr}) by $C_x$, we can now rewrite it as
\begin{equation}
\dot{\vec{Y}} = F\left(\vec Y\right) + \tilde C_{x}(t)/C_x
\label{eq:YeqFull}
\end{equation}
where  variable $\vec{Y}$ has been rescaled,
and $F\left(\vec Y\right)$ is the right-hand side of Eq. (\ref{eq:yrC}),
scaled by $C_x$.  Near the equilibrium solution of Eq.  (\ref{eq:yrC}),
Eq. (\ref{eq:YeqFull}) has the same form as Eq. (\ref{eq:averaging}), since both
$\tilde C_{x}(t)/C_x$ and $ F\left(\vec Y\right)$ are small.
Applying the 
Averaging Theorem (see for example \cite{guckenheimer}), we can thus
solve for $Y$,
\begin{equation}
Y = \bar Y + W
\label{eq:average}
\end{equation}
where $\bar Y$ is the solution of the averaged system, given by
Eqs. (\ref{eq:yrC}),  and $W$ is given by:
\begin{equation}
\frac{\partial W}{\partial t} = \tilde C_x
\label{eq:solution}
\end{equation}
Approximating $\tilde C_{x}$ to lowest order as a sinosoidal function at
frequency $N \omega$, we get an expression for $Y$ as a function of $\tilde
C_x$,
\begin{equation}
Y \approx \bar Y + \frac{\tilde C_x}{N \omega}
\label{eq:average2}
\end{equation}
Figure (\ref{fig:ave}) compares numerically calculated amplitude of $Y$ oscillation to the
one given by Eq. (\ref{eq:average2}), $\tilde C_x/N \omega$.  The
agreement is better at lower ultraharmonic amplitudes, in accordance to the
assumptions under which Eq. (\ref{eq:average2}) was derived: $\tilde
C_{x}(t)/C_x \ll 1$.  

As can be seen from Eq. (\ref{eq:average2}), the amplitude of ultraharmonic
oscillations is inversely proportional to the ultraharmonic frequency, $N
\omega$.  We thus expect a degradation of amplitude as the number of
oscillators in the array increases.  The next section proposes a different
coupling term that both achieves a higher amplitude of ultraharmonic
oscillations, and does not suffer degradation in amplitude as $N$ increases.  

\section{\label{sec:drive} V.  Coupling for achieving high 
frequency constant amplitude oscillations}
Up to this point, we have focused on 
amplitude dependent global coupling, of the form given by Eqs. (\ref{eq:xr})
and (\ref{eq:yr}).  
For this form of coupling, the amplitude of
ultra-harmonic oscillations is low and falls as $N$ increases. 
The amplitude of ultraharmonics cannot be significantly
increased by increasing the mutual coupling
strength, since high mutual coupling leads to a bifurcation of the $X$ array
and oscillator death.    
From the
previous sections, it should be 
clear that any form of coupling, which is a
periodic function of some frequency,  $\omega_f$, 
induces oscillations at that
frequency (for sufficiently low amplitude) as
long as the constant part of the coupling, $C_x$, is above the bifurcation
value:  $C_x > C_{bx}$.  It should therefore be 
possible to better control the amplitude of
ultra-harmonic oscillations and even the frequency
by the choice of the coupling function.  In this section, a form
of coupling is derived that induces ultraharmonic oscillations in the $Y$
array at frequency $2 \omega N$ (rather then $\omega N$), 
with an amplitude that stays relatively constant as $N$ increases. 
$N$ used will be odd, since as previously discussed, only $N$ odd arrays
have phases in increments of $2 \pi/N$ in the presence of diffusive coupling.  

\begin{figure}[ht]
\hspace*{-1 cm}
{
\epsfxsize=3.5in
\epsffile{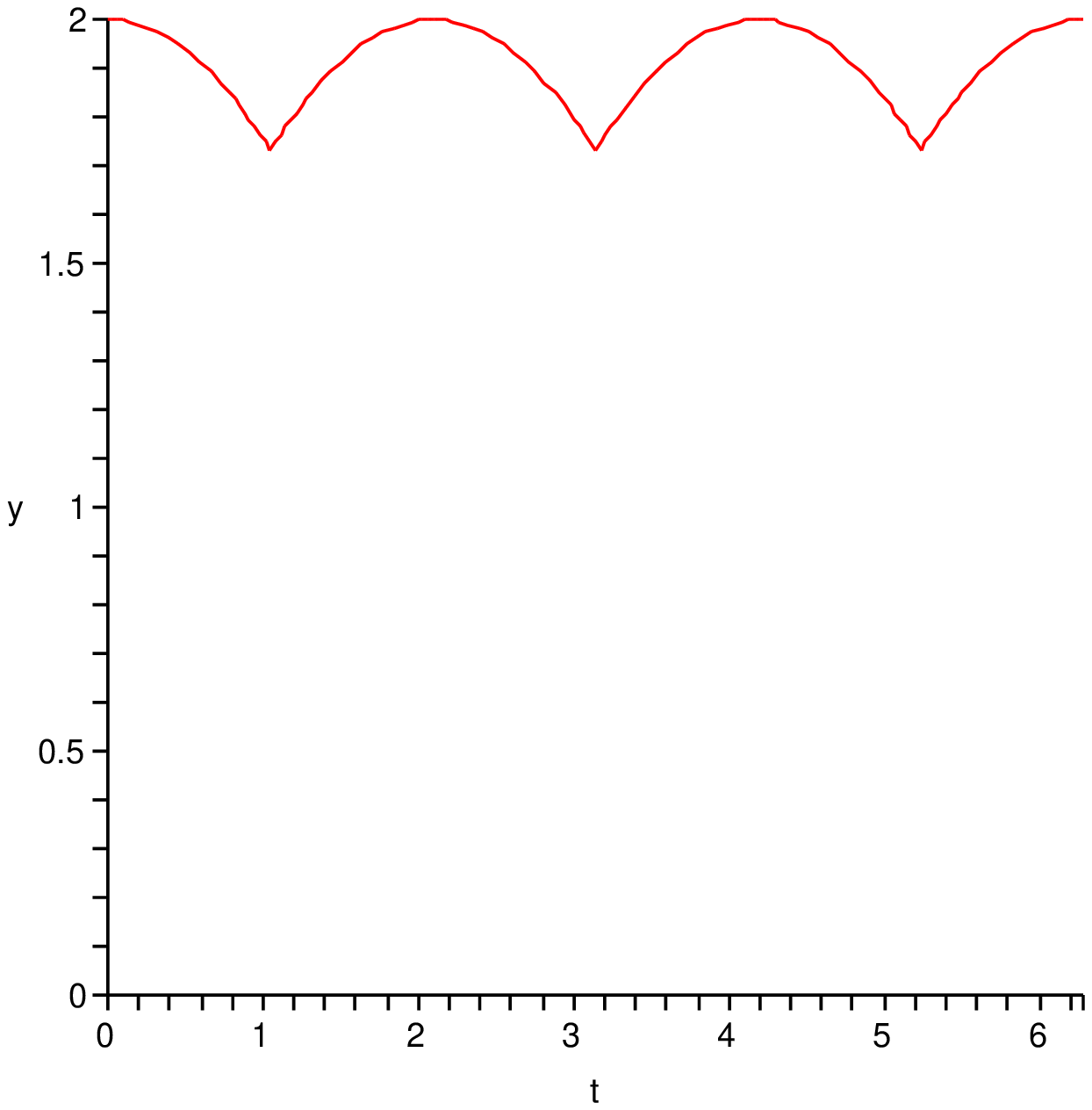}
}
{
\epsfxsize=3.5in
\epsffile{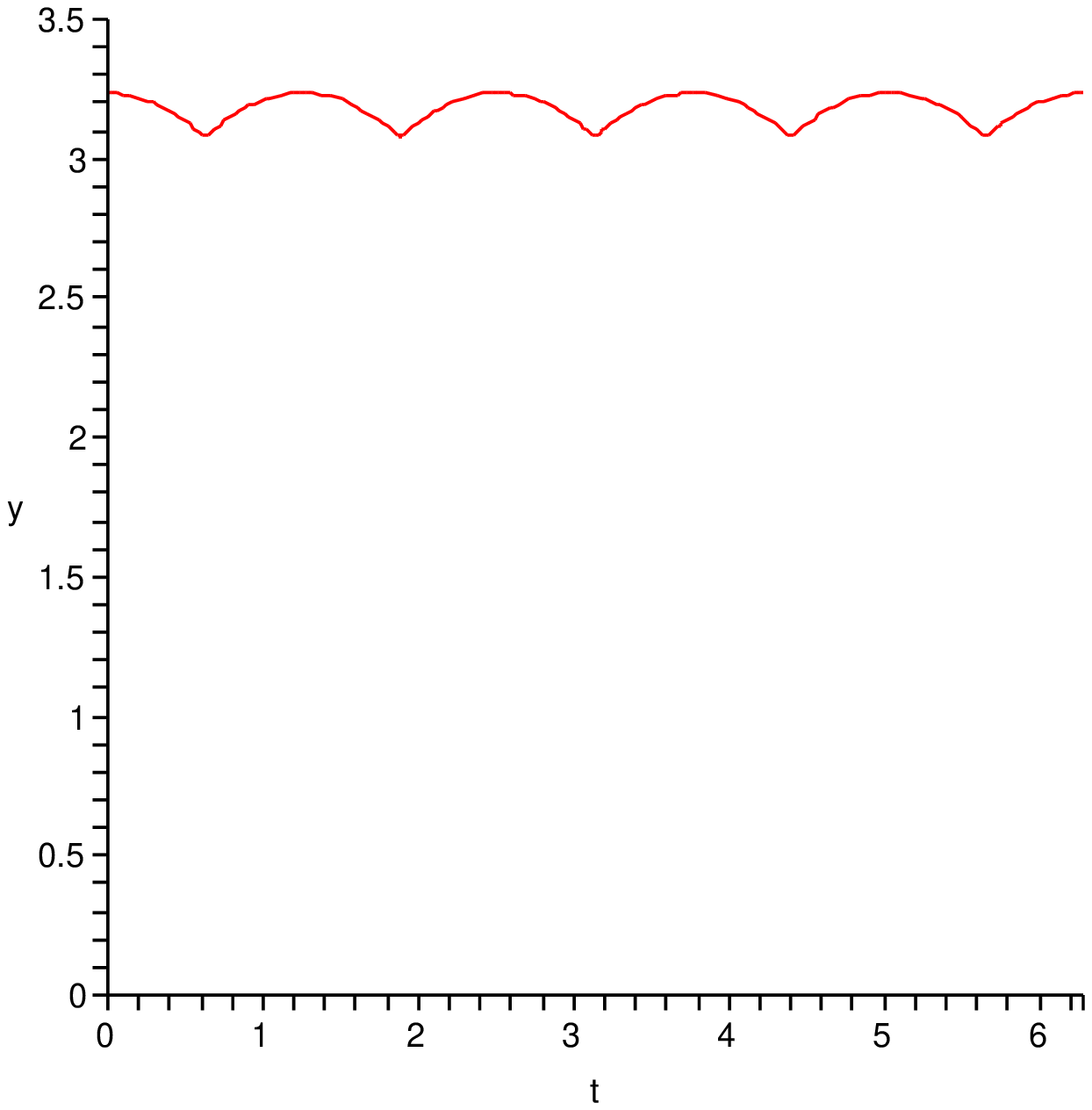}
}
\caption{\quad Top: $\sum_{j=1}^N |X_{jr}|$ for $N=3$. \quad Bottom: 
$\sum_{j=1}^N |X_{jr}|$ for $N=5$.  In both cases $\alpha=1$, $\omega=1/2$,
  $c_{yx} = 0$.  As $N$ increases, the constant component for this form of
  coupling 
  increases, while the relative value of the oscillatory component falls.}
\label{fig:9}
\end{figure}

As shown in Section II, 
in the absence of inter-array coupling, $c_{yx} =0$, 
the $X$ array oscillates as a collection of simple harmonic oscillators,
at an amplitude given by Eq. (\ref{eq:AmpOdd}), 
and with a phase difference of $\pi + \pi/N$ between nearest
neighbors and $2 \pi/N$ jumping over a neighbor.  
The function describing each oscillator in a steady-state is given by:
\begin{equation}
X_{jr} = A_x cos\left(\omega t - \phi_j\right)
\label{eq:Xmotion}
\end{equation}
where $X_{jr}$ denotes the real part of the $j$th oscillator, $\vec X_j$, and
$A_x$ is the
amplitude given by Eq. (\ref{eq:AmpOdd}).  The phase, $\phi_j$ is given by
\begin{equation}
\phi_j = \left(j-1\right) \left(\pi + \pi/N\right)
\label{eq:Phasej}
\end{equation}
where  Eq. (\ref{eq:Nodd}) was used.  
Since $\{X_{jr}\}$ is a collection of sinosoidal functions with phases
distributed in equal increments over the interval $[0, 2\pi]$, an
ultra-harmonic coupling function,  $c_{xy} \sum_{k=j}^N g(X_{j})$, 
can thus be created by simply summing the real part  of the
oscillation, $c_{xy} \sum_{k=1}^N g(X_{j}) = \sum_{j=1}^N |X_{jr}|$.  In this
case however, the constant component of the coupling, $C_x$ increases
substantially as $N$ increases, and the relative
amplitude of the ultra-harmonic
component, $\tilde C_x$ drops substantially with $N$ 
(where $\sum_{j=1}^N |X_{jr}| = C_x + \tilde C_x$).  Figure \ref{fig:9} shows
this effect for $N=3$ and $N=5$.  It is possible to
increase the amplitude of the oscillatory component of the coupling and
decrease $C_x$ by taking the sum of some power of $X_{jr}$,
\begin{equation}
c_{xy} \sum_{j=1}^N g(X_{j}) = c_{xy} N \sum_{j=1}^N |\tilde X_{jr}|^n
\label{eq:gencoupling}
\end{equation}
where $n$ is an integer yet to be determined, and $\tilde X_{jr}$ has been
normalized (divided by $A_x$):   
\begin{equation}
\tilde X_{jr} = cos\left(\omega t - \phi_j\right)
\label{eq:XmotionNorm}
\end{equation}
where $ \phi_j$ is given by Eq. (\ref{eq:Phasej}).
The factor of $N$ multiplying Eq. (\ref{eq:gencoupling}) is there to
compensate for the fall of ultraharmonic amplitude as the frequency increases
(see Section IV).  We need to find an expression for $n$ as a
function of $N$ and $C_x$, $n(N,C_x)$, 
such that the amplitude of oscillation stays fairly constant as
$N$ increases.   

\begin{figure}[ht]
\hspace*{-1 cm}
{
\epsfxsize=6in
\epsffile{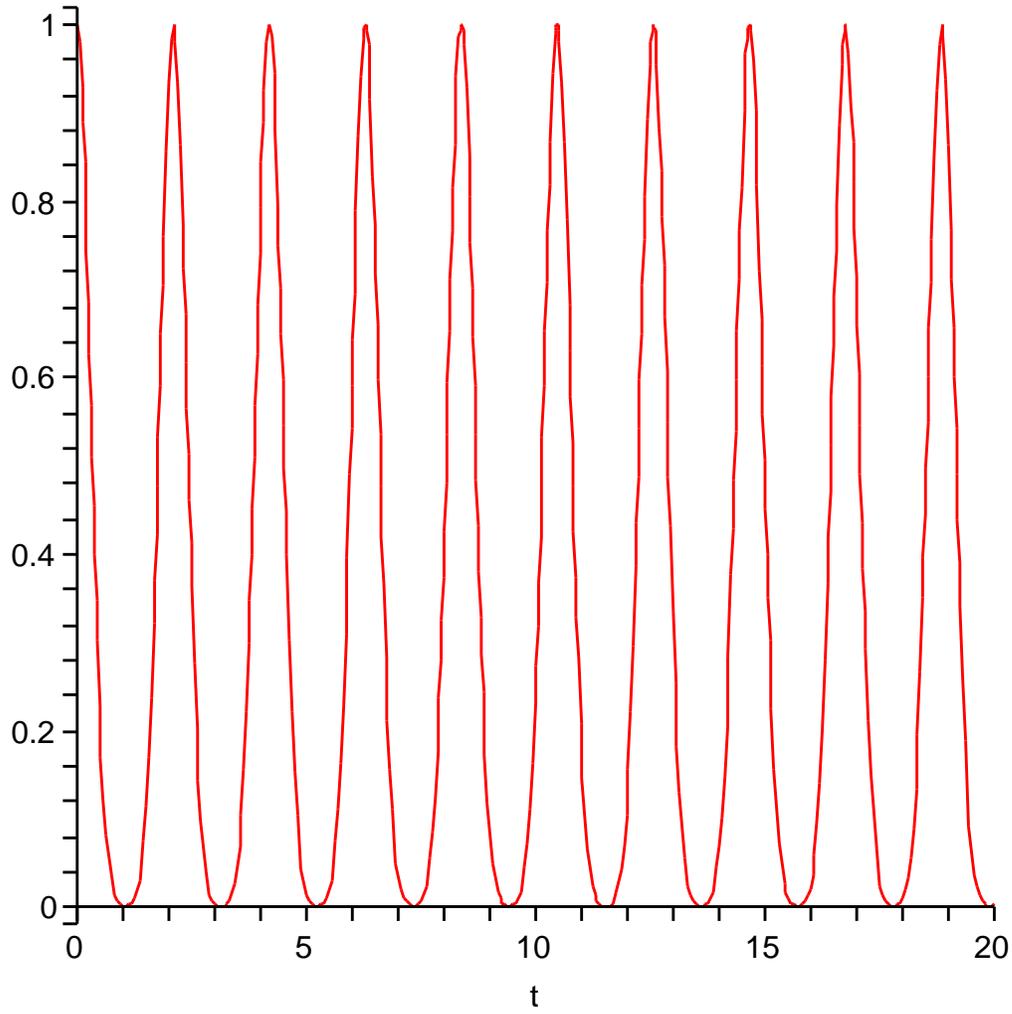}
}
\caption{Drive given by Eq. (\ref{eq:gencoupling}) for high n ($n=60$).  The peaks
  narrow substantially, with less overlap between neighboring peaks as $n$
  increases.}
\label{fig:10}
\end{figure}

\begin{figure}[ht]
\rotatebox{0}
{
\epsfxsize=3.5in
\epsffile{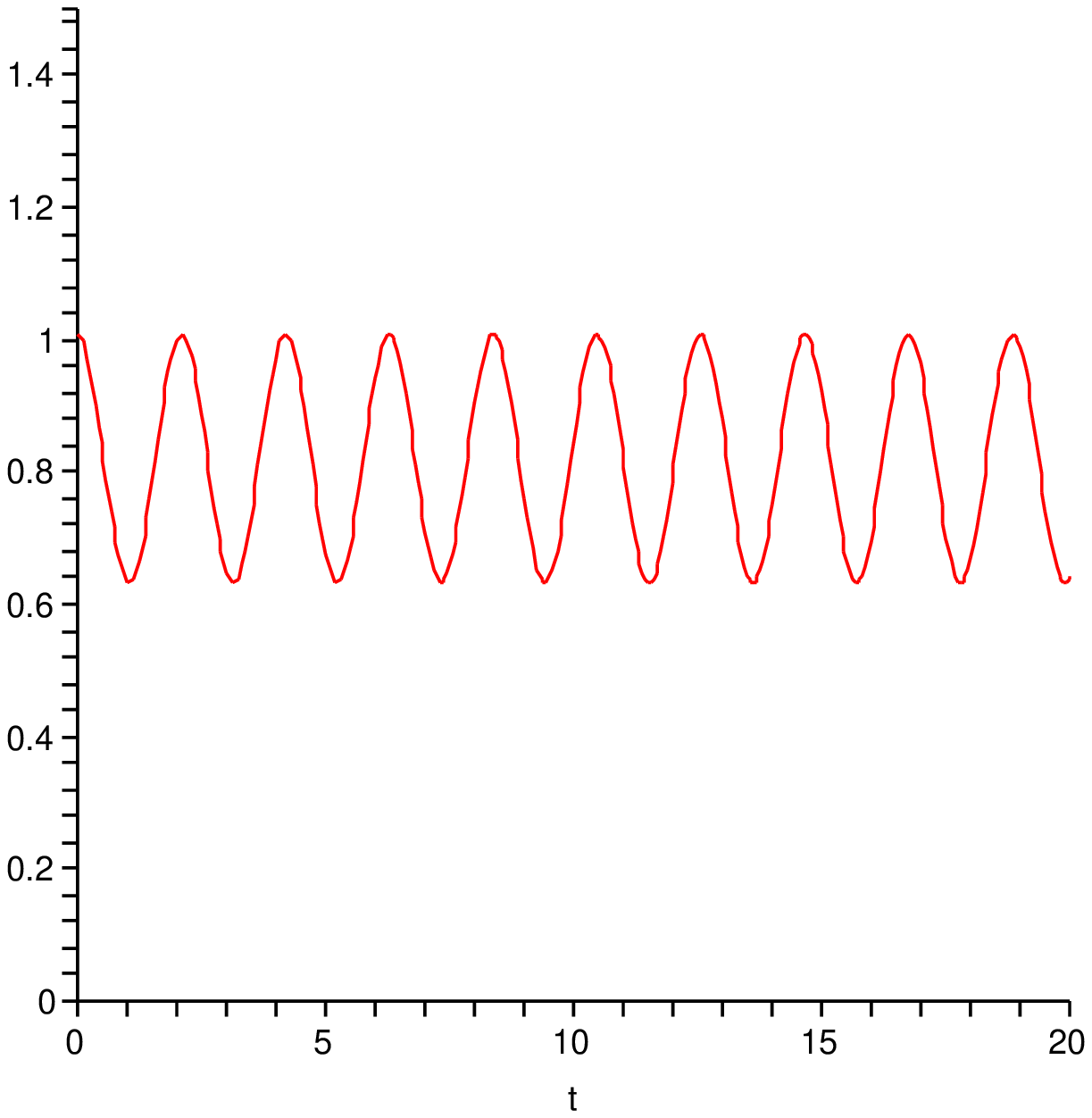}
}
\rotatebox{0}
{
\epsfxsize=3.5in
\epsffile{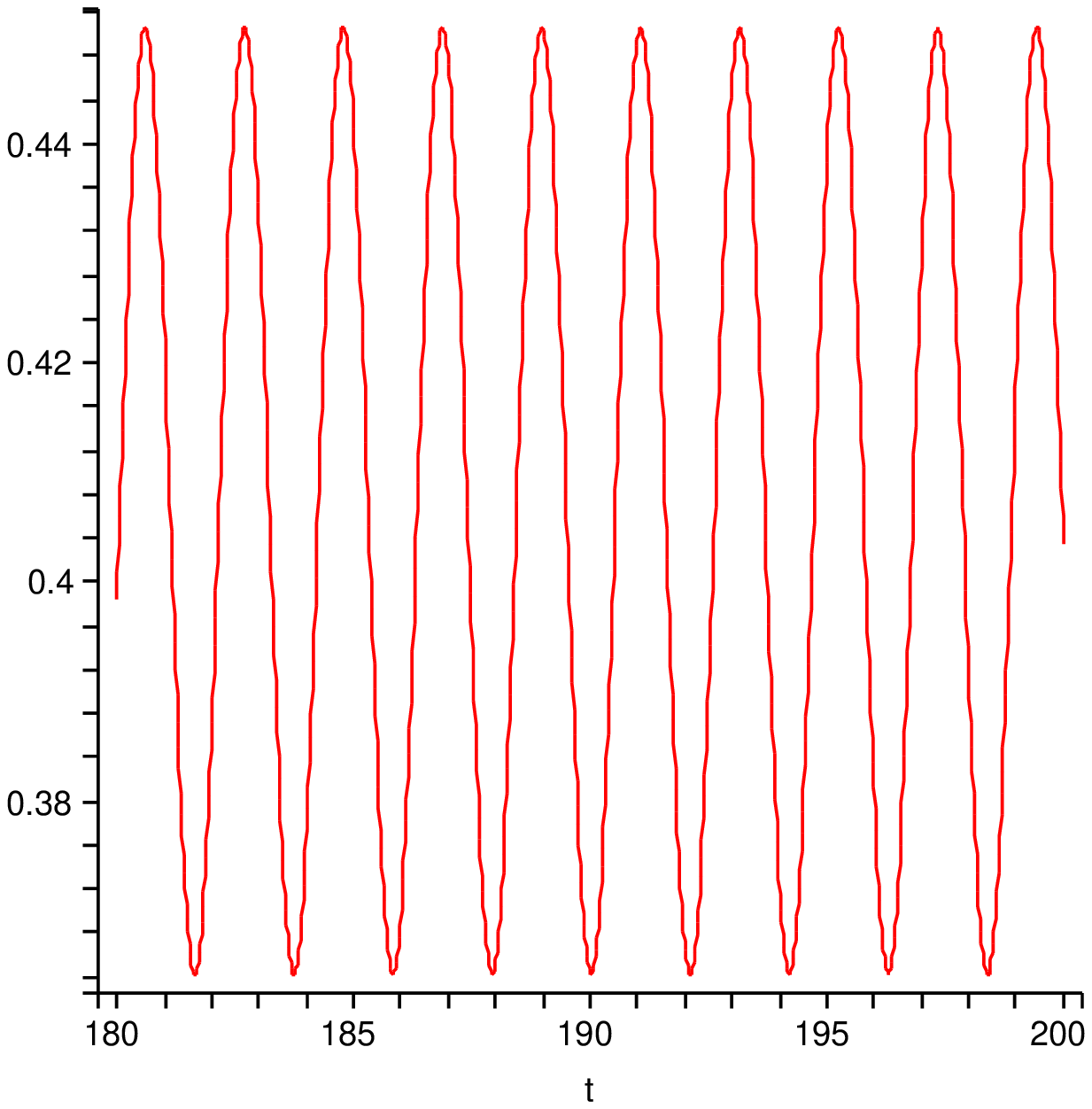}
}
\caption{Oscillations with coupling given by Eq. (\ref{eq:gencoupling}).
  Top: the coupling term, $c_{xy} N \sum_{j=1}^N |\tilde X_{jr}|^n$, with $C_x = 0.8$ and $\tilde C_x = 0.8$. 
 Bottom:  The ultraharmonic oscillations induced in the
  $Y$ array by the coupling shown at the top.  $\alpha=1$, $\omega=1/2$,
  $c_{xy} = 0.36$, $c_{yx}=0$, $N=3$.}
\label{fig:11}
\end{figure}

From Eq. (\ref{eq:XmotionNorm}), it is clear that increasing $n$ (for even
$n$) in Eq. (\ref{eq:gencoupling}) will lead to sharper, 
more narrow peaks centered
around $\omega t - \{\phi_j\} = 0, \pi$, 
with $j$ running between $1$ and $N$. 
Thus with increasing $n$, there will be less overlap between
neighboring peaks in Eq. (\ref{eq:XmotionNorm}), leading to a lower value of
$C_x$, the constant component of the coupling.
 (see Figure \ref{fig:10}).
In the limit as $n \rightarrow \infty$,
the sum in Eq. (\ref{eq:gencoupling}) becomes, as a function of $t$, 
\begin{equation}
c_{xy} \sum_{j=1}^N |\tilde X_{jr}|^n = c_{xy} \sum_{j=1}^N \int\limits^{t-1/n}_{t+1/n}
\delta\left(sin[\omega \tilde t - \phi_j]\right) d \tilde t
\label{eq:spikes}
\end{equation}
where $\delta$ is the delta function.
In steady state, and for $n$-even,  the normalized 
coupling from the
$X$ array as $n \rightarrow \infty$ 
is given by a series of equally-spaced
spikes, occuring at frequency $2 n \omega$ 
and of amplitude $c_{xy}$.  Figure \ref{fig:10} 
shows $\sum_{k=1}^N |\tilde X_{kr}|^n$
for $n=60$.  We can already see sharply defined spikes that approach a sum of
spikes of a unit amplitude as $n \rightarrow \infty$.
Eq. (\ref{eq:spikes}) has frequency of $ 2 N \omega$ rather then $N \omega$,
the frequency generated when the amplitudes are added.
This happens because  each $|X_{kr}|^n$ has 2 spikes over the time
interval $[0, 2 \pi/\omega]$, so that summing over $N$ oscillators leads to a
waveform of frequency $2 \omega N$, rather then $\omega N$, as for amplitude
coupled arrays  (see Figure \ref{fig:11}).  
Since neighboring peaks are separated by  $\pi/N$, the two nearest
spikes intersect at a phase difference of $\pi/2 N$ from the top of each
peak.  Assuming that the exponent, 
$n$, is sufficiently large so that only nearest neighboring
spikes have significant overlap (see Fig. \ref{fig:10}), we are led to the
following equation for the 
constant component of the coupling from the $X$-array to the $Y$-array:
\begin{equation}
C_x = 2 N \cdot c_{xy} [cos\left(\pi/2 N\right)]^n,
\label{eq:driveCx}
\end{equation}
where the exponent, $n$ is a function of $N$, 
such that $cos\left(\pi/2 N\right)$ stays constant as $N$ increases.
Therefore $C_x$ increases linearly with $N$, a situation similar
to the amplitude coupled arrays, where the constant component of the coupling
also increases linearly as $N$, for large $N$ (see Eq. (\ref{eq:cxSimple})).
The amplitude of the oscillatory component of the coupling is found
by subtracting the constant component, $C_x$, 
given by Eq. (\ref{eq:driveCx}) from the peak of the normalized oscillation,
which occurs at one,
\begin{equation}
max(\tilde C_x) = c_{xy} N \left(1- 2 [cos\left(\pi/2 N\right)]^n\right).
\label{eq:A_xc}
\end{equation}
Thus, unlike amplitude coupled arrays, the oscillatory component of the
coupling also increases linearly with $N$, which prevents the degradation of
amplitude seen in the amplitude coupled case.
Using Eq. (\ref{eq:average2}) with $2 N \omega$ in the denominator (since that
is the frequency of the drive), the amplitude of ultraharmonic oscillations can
be approximated as 
\begin{equation}
A_y \approx \frac{c_{xy}}{2 \omega} 
\left(1- 2 [cos\left(\pi/2 N\right)]^n\right)
\label{eq:DrivenAmp}
\end{equation}
and should therefore stay fairly constant as $N$ 
increases if $[cos\left(\pi/2 N\right)]^n \sim const$ for large $N$.  
For large $N$, we can approximate the cosine term as
\begin{equation}
cos(\pi/2 N) \approx 1 - (\pi/2 N)^2 \approx 
exp\left(- \frac{\pi^2}{2 N^2}\right)
\label{eq:cosApprox}
\end{equation}
where the first approximation came from taking the first two terms of 
a series expansion of a cosine
function and the 
second approximation came from expanding the exponential in a
series, taking the first two terms and comparing them to $1 - (\pi/2 N)^2$.  
Thus in order for $A_y$ in Eq. (\ref{eq:DrivenAmp}) to stay fairly
constant as $N$ increases, for $N$ large,
we need $n \propto N^2$.  When taking the power, $n$ should be rounded
to the nearest even number.  Choosing 
\begin{equation}
n = K N^2 - K N^2 mod 2 \qquad if \qquad K N^2 mod 2 \leq 1
\label{eq:n1}
\end{equation}
\begin{equation}
n = K N^2 - K N^2 mod 2 + 2 \qquad if \qquad K N^2 mod 2 > 1.
\label{eq:n2}
\end{equation}
The above equations
ensure that $n$ is rounded to the nearest even number.
$K$ is some constant,
 which can be chosen to achieve a desired value of
$C_x$  (for example $K=2/3$ is a good choice).
Using Eqs.  (\ref{eq:DrivenAmp})-(\ref{eq:n2}) 
we obtain an approximate
expression for the amplitude of ultraharmonic
oscillations when $N$ is large: 
\begin{equation}
A_y \approx \frac{c_{xy}}{2 \omega} 
\left(1- 2 exp\left(- \frac{\pi^2 K}{2}\right)\right).
\label{eq:DrivenAmpSol}
\end{equation}
From the above equation, the coupling strength, $c_{xy}$ and $K$ can be chosen
to obtain the desired amplitude of ultraharmonics, with a maximum possible
amplitude being $A_y \approx c_{xy}/2 \omega$.  

\begin{figure}[ht]
\hspace*{-1 cm}
{
\epsfxsize=7.5in
\epsffile{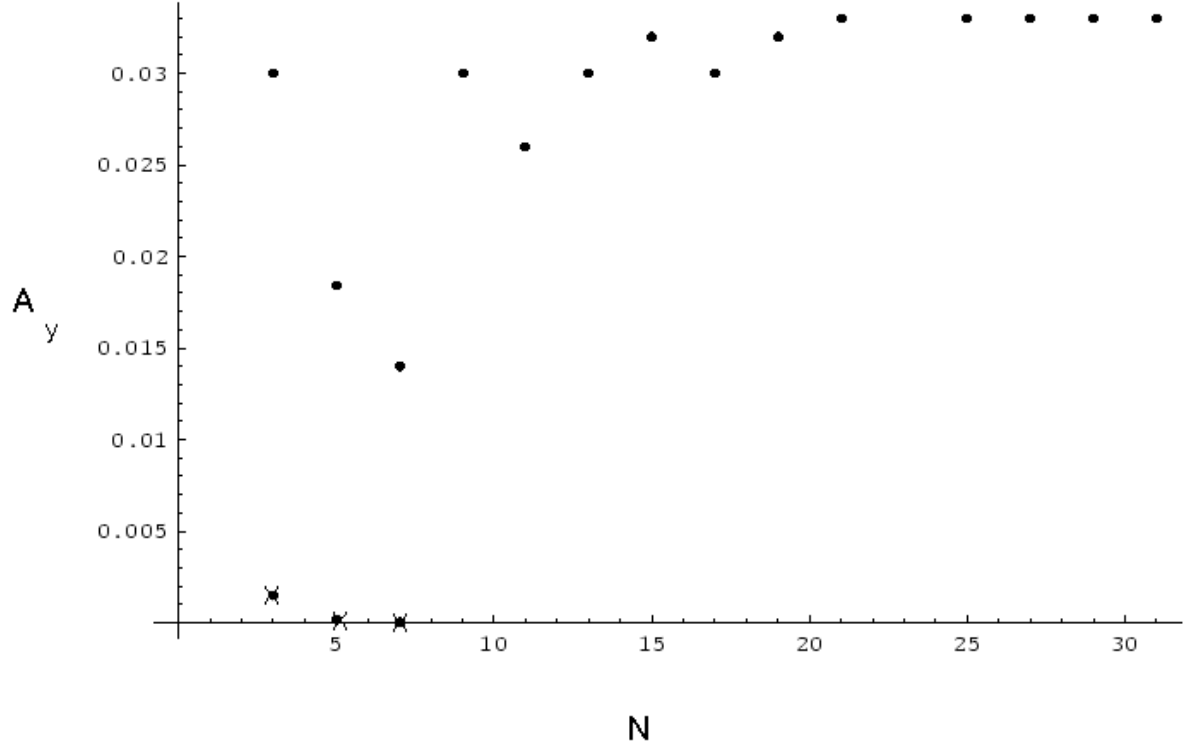}
}
\caption{Amplitude of ultraharmonic oscillation, $A_y$, with the drive
coupling, as a function of $N$, $K=2/3$.  
The lower three points, marked by crosses,  correspond
to amplitude of ultraharmonics with the standard
amplitude coupling.  The constant part of the coupling, $C_x$ is the same for
both cases, $c_{xy}=c_{yx}=.1134$, $\alpha_x=\alpha_y=1$, $\omega=1/2$}
\label{fig:drive}
\end{figure}

Figure (\ref{fig:drive}) shows the amplitude of ultraharmonics, $A_y$, as a
function of $N$ for both the drive coupling given in
Eq. (\ref{eq:gencoupling}), with $n$ given by Eqs.  (\ref{eq:n1}) and
(\ref{eq:n2}) and mutual coupling analyzed in the previous section.  The
parameters were chosen such that the constant component of the coupling, $C_x$
is the same in both cases.  The solid line for the drive dependent coupling is
the result of connecting the data points for the odd $N$ values between $N=3$
and $N=31$.  The amplitude of ultraharmonics asymptotes to a constant value of
around $0.033$, close to the value 
predicted by using Eq. (\ref{eq:DrivenAmpSol})
of around $0.028$.  It is clear that the amplitude of ultraharmonics generated
by the drive is significatly higher than that by the mutual coupling, given by
Eqs. (\ref{eq:xr}) and (\ref{eq:yr}).  In addition, the drive coupling does
not suffer from oscillator death, thus the amplitude will stay constant while
$N$ increases.  In contrast, increasing $N$ for mutual coupling leads to
oscillator death, unless the coupling constants $c_{xy}, c_{yx}$ are also
decreased (leading to an even further fall in amplitude).  In Figure
(\ref{fig:drive}), oscillator death occurs at just $N=9$, for the coupling
used.  
Thus the drive has many advantages,
such as a relatively high, easily controllable amplitude that does not degrade
with an increase in $N$ and does not suffer from oscillator death, which
occurs in mutually coupled arrays.  However, since $n$ increases as $N^2$,
high ultraharmonics, $N \omega$, require rather high powers in the coupling
function.  

\section{\label{sec:conclusion} VI. Conclusion}

The mechanism behind the generation of ultra-harmonic oscillations in two
mutually coupled arrays of limit cycle oscillators was analyzed.
These ultra-harmonic oscillations were shown to occur as a result of a
bifurcation that results when the coupling from the $X$ to the $Y$ array
exceeds a certain value.  This coupling consists of a large constant
component, since the coupling is amplitude-dependent, and a smaller
ultraharmonic oscillatory component
that results from the breaking of symmetry in the
$X$-array due to coupling from the $Y$-array. 
This smaller oscillatory component of the coupling induces the $Y$ array to
oscillate at an ultra-harmonic frequency around the equilibrium determined by
the large constant component of the coupling.  It was also shown that in the
case of amplitude-dependent coupling, the ultra-harmonic oscillation is the
result of a mutual interaction between the two arrays (rather than a
master-slave system), since the coupling from the $Y$ to the $X$ array is
necessary to induce an oscillation in the otherwise conserved amplitude of $X$
oscillators.  

The range of inter-array coupling constants whereby
ultra-harmonic oscillations are produced was derived for symmetric
$c_{xy}=c_{yx}$ coupling, but can be generalized to non-symmetric
coupling, where $c_{xy} \neq c_{yx}$.  For symmetric coupling, this range
depends on the limit cycle frequency, the amplitude of oscillation and the
strength of nearest-neighbor coupling within the $X$ array.  
The allowable range of
inter-array coupling constant increases with an increase in strength of the
diffusive coupling, $c_x$, in the out-of-phase coupled $X$ array.  
Thus higher
absolute values of $c_x$ lead to a greater possible range of values of the
inter-array coupling constant whereby ultra-harmonic oscillations are created.
The $Y$ array diffusive coupling strength, $c_y$, on the other hand, 
does not affect the conditions for creation of stable ultraharmonics,
since the diffusive coupling term drops out in steady-state
for in-phase coupled arrays.  
The derived bifurcation values for
the inter-array coupling constant agree well with numerical simulation, and
can be used to tune the value of the coupling constant to control the
amplitude of the ultra-harmonic oscillations.   

For achieving
better-controlled, higher amplitude ultra-harmonic oscillations,
another form of coupling was suggested.
This one-way coupling has the advantage of achieving 
higher amplitude ultra-harmonic
oscillations that do not fall in amplitude as the number of oscillators, and
therefore the ultra-harmonic frequency, increases.  It also does not suffer
from oscillator death, which puts an upper limit on the strength of symmetric
coupling that can be used in amplitude coupled arrays.  The suggested form of
coupling, however,  requires increasingly 
more complicated forms
of the coupling function as $N$ increases
and may be more difficult to implement experimentally.

Finally, though the methods of analysis here were applied to
oscillator of Stuart-Landau type, they may be applied to various
applications of interest, where both frequency and power control are
required. Such examples of  the stabilization of in-phase arrays occur
in such areas 
as electronic circuits for radar \cite{InKNPLM03} , phase locked
nano-scale magnets used for microwave sources \cite{MancoffRET05}, 
power systems \cite{AbedV84}, and Josephson junction arrays used for
terahertz sources \cite{AronsonGK91}.

\section{Acknowledgements}
This work was supported by a grant from the Office of Naval
Research.  ASL
is currently a post doctoral fellow with the National
Research Council.


\eject

%
%

\end{document}